\documentclass[a4paper,11pt]{elsarticle}
\usepackage{jinstpub} 
\usepackage{subfigure}
\usepackage{graphicx}
\usepackage{soul}
\usepackage{lineno}
\usepackage{amssymb}
\usepackage[colorlinks=true]{hyperref} 

\begin{document}

\title{Characterization of Hamamatsu R11410-23 Photomultiplier Tubes and Performance in the PandaX-4T Detector}

\author[a,1]{Binbin Yan, \note{Corresponding author. yanbinbin@sjtu.edu.cn.  On behalf of all authors, the corresponding author states that there is no conflict of interest.}}
\author[b]{Di Huang,}
\author[a,b]{Jianglai Liu,}
\author[c,d]{Xiaoying Lu,}
\author[c,d]{Anqing Wang,}
\author[b]{Mengjiao Xiao.}

\affiliation[a]{Tsung-Dao Lee Institute, Shanghai Jiao Tong University, Shanghai, 200240, China}
\affiliation[b]{School of Physics and Astronomy, Shanghai Jiao Tong University, Key Laboratory for Particle Astrophysics and Cosmology (MoE), Shanghai Key Laboratory for Particle Physics and Cosmology, Shanghai 200240, China}
\affiliation[c]{Research Center for Particle Science and Technology, Institute of Frontier and Interdisciplinary Science, Shandong University, Qingdao 266237, Shandong, China}
\affiliation[d]{Key Laboratory of Particle Physics and Particle Irradiation of Ministry of Education, Shandong University, Qingdao 266237, Shandong, China}


\abstract{ 
The PandaX-4T liquid xenon detector uses Hamamatsu 3-inch R11410-23 photomultiplier tubes (PMTs) as the light sensors for ultra-low radioactivity, high quantum efficiency, and long-term stability of all PMTs at cryogenic temperature. 
Each PMT was thoroughly tested in a dedicated chamber before being installed in the detector to ensure compliance with experimental requirements.
Main PMT characteristics, including gain, dark count rate, and after-pulse probability, were measured and the distributions of these parameters were presented.
Additionally, all PMTs were tested in a cryogenic environment to simulate their operating conditions in an actual detector environment.
Finally, we report the long-term performance of all PMTs during the commissioning run of the PandaX-4T experiment, and most of the PMTs worked well.}

\date{\today}
\maketitle

\section{Introduction}
\label{sec:intro}
Dual-phase xenon time projection chambers (TPCs), such as those from PandaX-4T~\cite{PandaX-4T:2021bab}, XENONnT~\cite{XENON:2020kmp}, and LZ~\cite{LZ:2021xov} are important players in the dark matter direct detection and neutrino physics.
The detectors are leading the search for weakly interacting massive particles (WIMPs), a leading theoretical candidate for dark matter~\cite{Bertone:2004pz,Jungman:1995df}, with masses ranging from a few GeV/$c^2$ to several TeV/$c^2$\cite{LZ:2022lsv,XENON:2023cxc,PandaX:2024qfu}.
The experiments increasingly emphasize the application of xenon detectors in neutrino physics, especially in neutrinoless double beta decay and solar neutrino searches~\cite{PandaX:2022kwg,PandaX:2023ggs,PandaX:2024muv,PandaX:2024jjs}.

In these detectors, a physical event produces a prompt signal (S1) from scintillation photons in the liquid xenon and a delayed signal (S2) from ionized electrons that drift to the liquid surface and produce electroluminescence photons in the xenon gas.
Both signals are recorded by photomultiplier tubes (PMTs) on the top and bottom of the detector. 
For the current generation of experiments, Hamamatsu R11410 3-inch PMTs remain the preferred light sensor for the 175-nm vacuum ultraviolet (VUV) photons emitted by xenon.
Hamamatsu R11410 PMTs have low radioactivity, low dark count rate (DCR), high quantum efficiency (QE), and high single-photon detection efficiency.

These PMTs have been used since the beginning of the PandaX experiment.
In PandaX-I experiment~\cite{PandaX:2015gpz}, we selected R11410-MOD/10 PMTs with high QEs to potentially achieve a relatively larger S1 signal to emphasize low-mass WIMP detection~\cite{Aalbers:2022dzr}. 
However, PandaX-I observed a high accidental coincidence background due to a high DCR. 
Additionally, the presence of $^{60}$Co and $^{110m}$Ag in the Kovar and other raw materials increased radioactivity.
In PandaX-II~\cite{PandaX-II:2020oim}, the focus shifted toward achieving lower dark rates and radioactivity with a new iteration of PMTs, Hamamatsu R11410-20. 
The PMT leakage in liquid xenon environments, which causes a high after-pulse, also has to be addressed. 

PandaX-4T, the current generation of PandaX, uses the new Hamamatsu R11410-23 PMTs for light readout. 
All PMTs were checked for intrinsic radioactivity using an underground high-purity germanium (HPGe) counting station. 
The measured average radioactivity levels of ${^{238}}$U and ${^{232}}$Th per PMT were below 3.99~mBq and 3.06~mBq~\cite{PandaX-4T:2021lbm}, respectively.
We selected the 368 best-performing units, prioritizing lower dark count rate (DCR), lower after-pulse probability, and better operational stability in our offline characterization process, with all other qualified PMTs reserved as backups. 
With those PMTs, PandaX-4T successfully commissioned from November 28, 2020, to April 16, 2021, collected and published 95 calendar days of stable physics data~\cite{PandaX-4T:2021bab, PandaX:2022osq,PandaX:2024pme,PandaX:2023toi}. 

The remainder of this paper is organized as follows.
Section~\ref{sec:setup} discusses the general test methodologies and experimental setup.
Section~\ref{sec:commissioning run performance} describes the PMT assemblies in PandaX-4T and long-term stability results.
Section~\ref{sec:summary} concludes the paper.

\section{Testing procedure and PMT characterization}
\label{sec:setup}
\begin{figure}[htbp]
  \centering
  \includegraphics[width=0.7\textwidth]{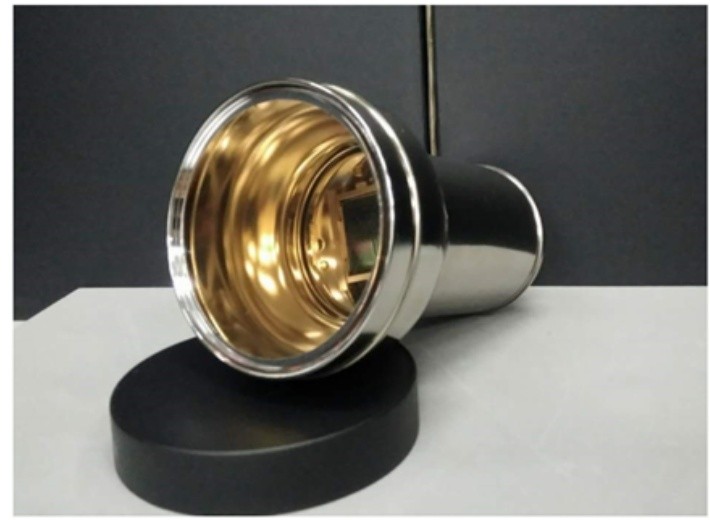}
  \caption{A photograph of R11410 3~inch PMT.}
  \label{fig:pmt photo}
\end{figure}

The 3-inch Hamamatsu R11410-23 PMT was specifically designed for low background and UV light applications, such as large underground xenon detectors.
It uses a bialkali photocathode with an area of 41~cm$^2$, achieving a quantum efficiency of 34\% for scintillation light from liquid xenon at the characteristic wavelength of 175~nm.
The PMT has 12 stage box and a linear-focused style dynode structure, which enables efficient electron multiplication.
The dynodes and photocathode are enclosed in a kovar metal body support
structure with a ceramic stem and kovar leads shown in figure~\ref{fig:pmt photo}.

All PMTs used in the PandaX-4T detector underwent extensive testing.
We applied a high voltage of approximately 1500V with two polarities and used three warm and cold circles to check stability.
Each PMT was tested in nitrogen at room temperature and -90\,$^\circ$C. The testing facility is described in Section~\ref{sec:tesing facility}. The main parameters, such as gain, single-photon-resolution, DCR, after-pulse probability (APP), and light emission, are introduced in Section~\ref{sec:offline parameters}.
Any PMT exhibiting out-of-range parameters during this procedure was rejected.

\subsection{Testing facility}
\label{sec:tesing facility}
\begin{figure}[htbp]
  \centering
  \includegraphics[width=0.96\textwidth]{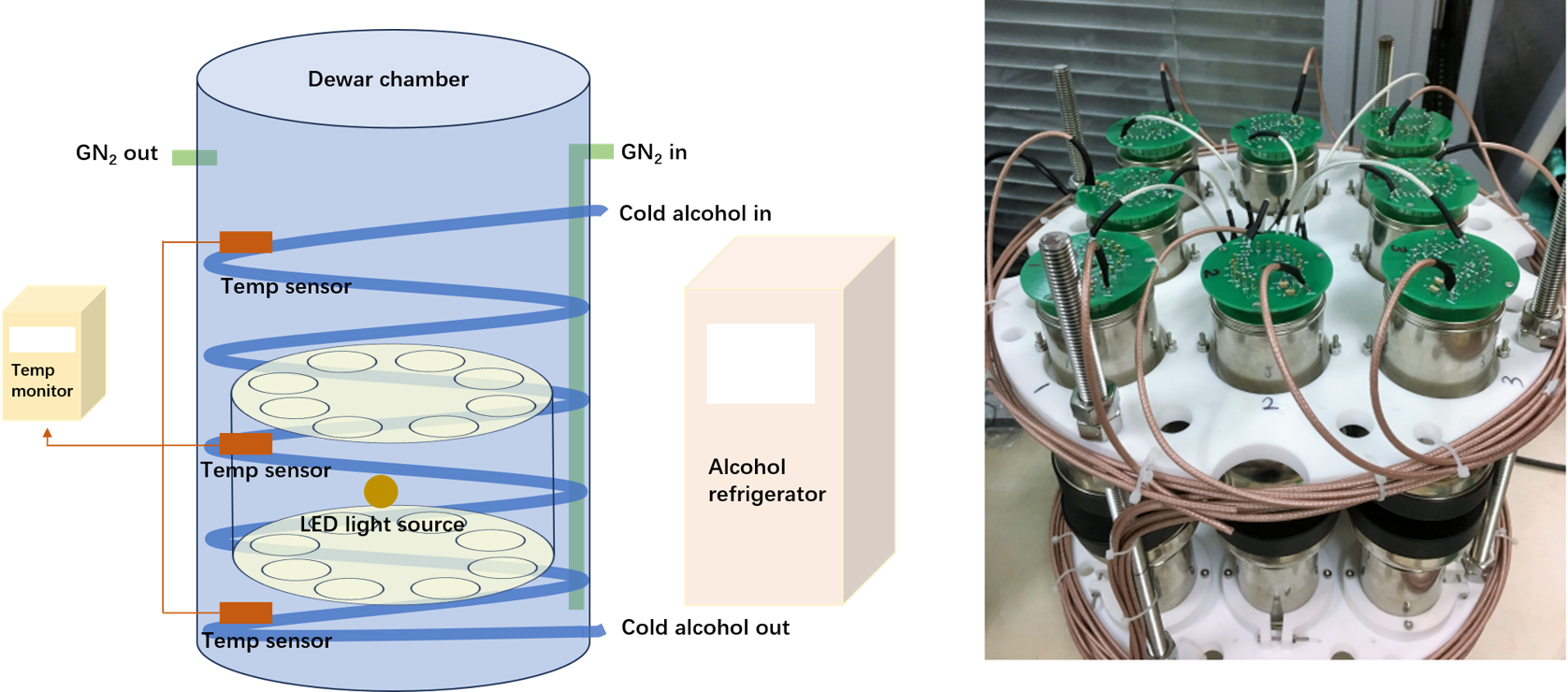}
  \caption{Schematic(left) and photo(right) of PMTs' support structure: 16 PMTs were tested simultaneously. In the right figure, the white cables are connected to a negative high-voltage distribution plate; the pink cables are connected to positive high-voltage sources.}
  \label{fig:pmt support structure}
\end{figure}

As shown in Fig~\ref{fig:pmt support structure}(left), the overview schematic diagram exhibits a setup incorporating a Dewar chamber, a segment of cold alcohol tube, three temperature sensors, and an LED light source. 
For temperature regulation, an alcohol refrigerator was utilized, enabling the chamber temperature to vary from room temperature to -90~$^\circ$C, which is slightly higher than the liquid xenon temperature.
Three temperature sensors were placed at the chamber's top, middle, and bottom to monitor the temperature. 
Sixteen PMTs were placed inside the chamber in a two-layer polytetrafluoroethylene (PTFE) support structure with eight PMTs on each layer, facing each other, as shown in Fig.~\ref{fig:pmt support structure}(right). 
At the beginning of the testing procedure, a room-temperature test was conducted.
Subsequently, the Dewar flask was cooled from room temperature to -90~$^\circ$C over 6~h and maintained at this temperature for 12~h. 
The PMT characteristic parameters were monitored hourly during the cryogenic phase. 
Upon completion of the cryogenic test, the Dewar was warmed to room temperature. 
Three cycles were performed from room temperature to the cryogenic temperature and back to room temperature.

Before each test, the Dewar chamber was flushed with dry nitrogen gas for at least 10 minutes. 
Subsequently, the chamber was pressurized with nitrogen (3~bar) at room temperature.
As the temperature decreased, the pressure in the chamber remained positive throughout the process. 
The feedthrough was sealed with epoxy glue to avoid the sparking from the condensed water at low temperatures.
A 375~nm light-emitting diode (LED) driven by a waveform generator was used to generate photoelectrons from the PMT photocathode, for the measurement of core parameters including gain and after-pulse probability. These parameters have negligible dependence on incident light wavelength, and the PMT response to the 175 nm liquid xenon scintillation light is calibrated in-situ in the PandaX-4T detector.
The light from the LED passed through an optical fiber into the chamber and was diffused by a 1~cm diameter PTFE ball.
A synchronous trigger was provided for the data acquisition (DAQ) system simultaneously with the LED light emission. 
This trigger mode, called an external trigger, was later used to characterize the PMT parameters.

\begin{figure}[htbp]
\centering 
\subfigure{
    \label{fig:circuit diagram}
    \centering
    \includegraphics[width=0.9\textwidth]{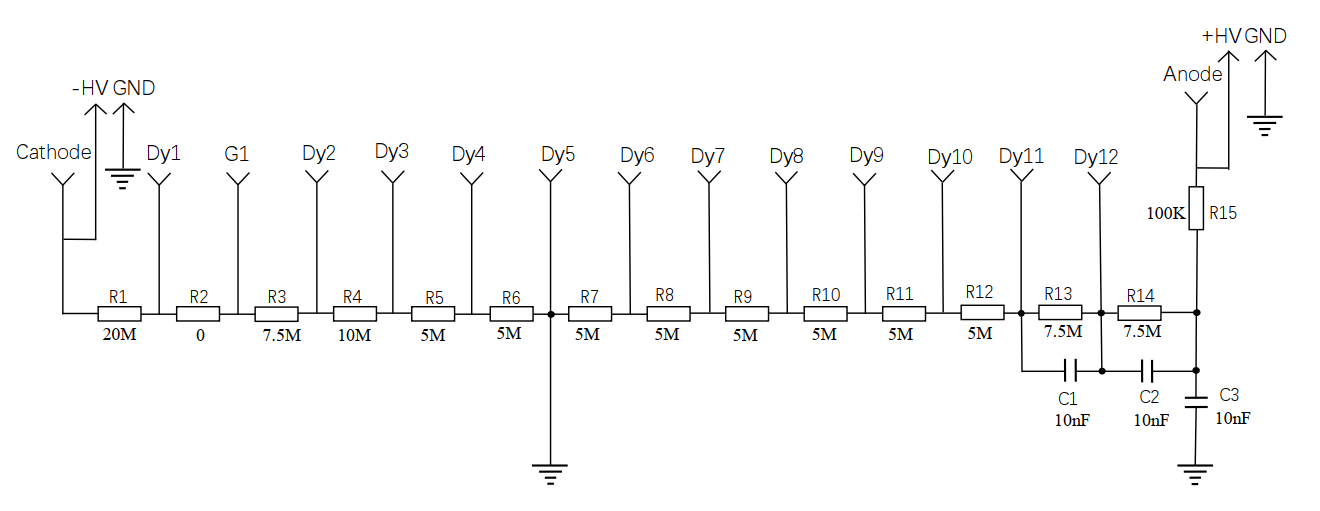}
}
\subfigure{
    \label{fig:base_photo}
    \centering
    \includegraphics[width=0.85\textwidth]{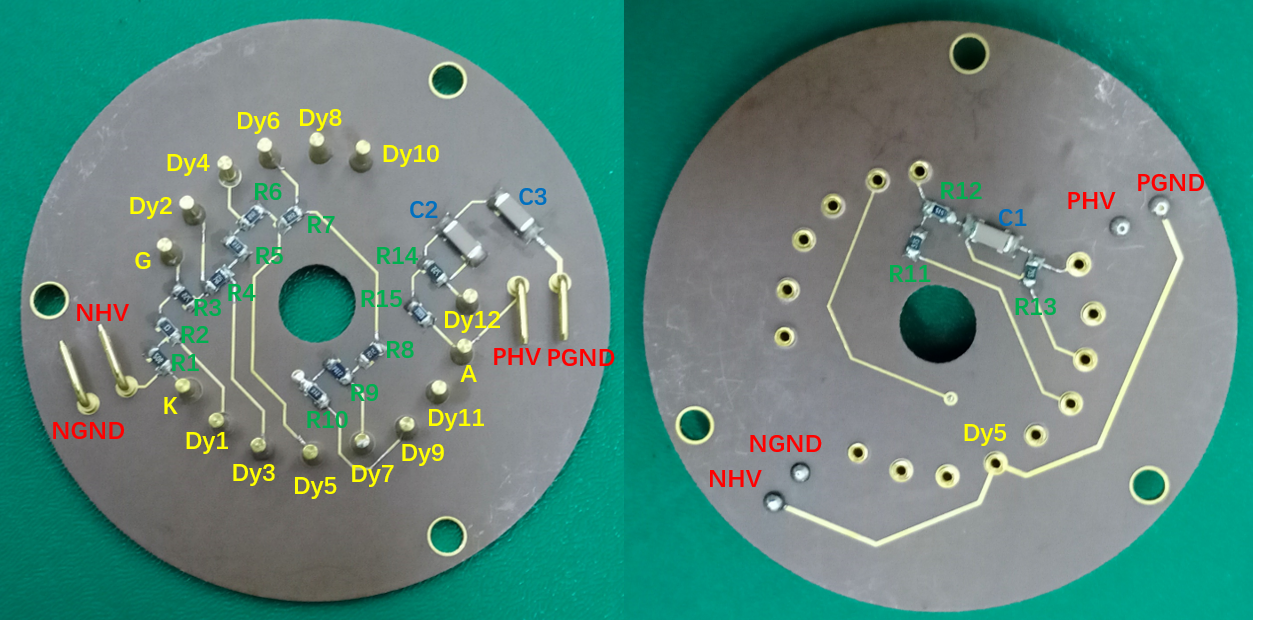}
}
\caption{(upper) schematic diagram of High-voltage divider base circuit. (lower) Photo of high-voltage divide base for PMT test. The dynode pins, resistors, capacitors, and high-voltage source elbows are marked in yellow, green, blue, and red, respectively. }
\label{fig:base}
\end{figure}

A customized low-radioactivity high-voltage divider base, shown in Fig.~\ref{fig:base}, was designed for WIMP detection in low-energy regions according to Hamamatsu's recommended voltage distribution ratio. 
The resistance chain is 92.5~M$\Omega$, which led to a negligible thermal power of 0.02~W at 1500~V voltage. 
Two capacitors labeled C2 and C3 in the figure, were used to maintain the voltage between the Dy11-Dy12-anode whenever PMT received a large signal. 

Another notable feature of the base is the bipolar biasing~\cite{Elsied:2015ixa}.
The base was designed with dual high-voltage input ports grounded at the fifth dynode.
For typical PMT operations, approximately 750~V negative high voltage (NHV) and positive high voltage (PHV) were applied to the cathode and anode, respectively.
The bipolar biasing reduced the voltage on the HV cables and lowered the requirements for voltage rating on cables and feedthroughs.
It minimized sparking between the components on the base and at the feedthrough pins. 
The PHV cable carried the PMT signal, and the direct current (DC) component was removed in a de-coupler box before the DAQ system. 
To simplify cabling and the number of feedthroughs, eight PMTs shared one NHV cable.

\begin{figure}[htbp]
  \centering
  \includegraphics[width=0.98\textwidth]{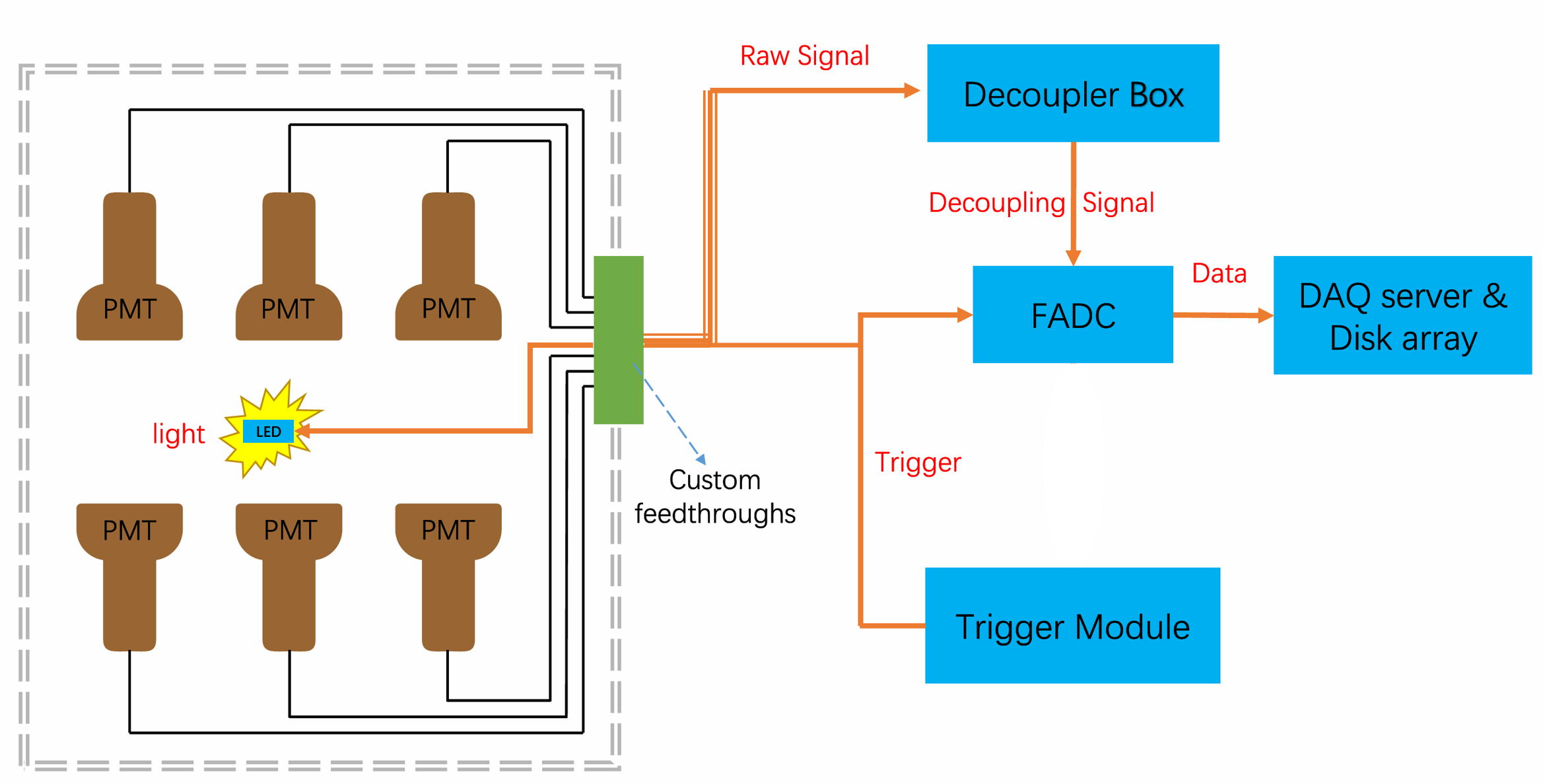}
  \caption{schematic diagram of the connection structure from PMT to the servers.}
  \label{fig:pmt assemblies and read out circuit}
\end{figure}

The signals were digitized in a flash analog-to-digital converter (FADC) from CAEN (model number V1725B, sampling rate 250M) with 16 channel inputs and 14-bit resolution. 
The dynamic range was configured to 2.0~V peak to peak, leading to a ratio of 0.122~mV per ADC unit~\cite{Yang:2021hnn}.
The readout system, shown in Fig.~\ref{fig:pmt assemblies and read out circuit}, can operate in the global self-triggered mode, which means the FADC board digitizes the waveform formed by all 16 channels if one of the channels receives a signal above the threshold.
The trigger threshold was set at 20~ADC (2.4~mV) unit, which responds to approximately one-third the height of a typical signal from single-photon electrons (SPEs).
The fixed 20~ADC count threshold for DCR calculation primarily suppresses baseline and electronic noise, excluding non-physical fluctuations; notably, all PMTs will be modified to 5$\times$10$^6$ gain during PandaX-4T operation, while SPE efficiency variations across different thresholds are calibrated via LED measurements and incorporated into waveform simulations for correction~\cite{PandaX:2024med}.
The data were stored on a server for subsequent analysis.

The signal was read out using a 48-pin Kyocera custom feedthrough and two LEMO cables, similar to the PandaX-4T readout system.
The LEMO cable between the inner and outer vessels is 10~m, and from the outer vessel to the DAQ system, it is 5~m.
The cross-talk effect in the entire readout chain was evaluated by calculating the charge ratio between the triggered channel and the recorded waveform of each channel in the global self-triggered mode. 
All feedthroughs and cables used for PMT signal readout were tested, the cross-talk ratio is less than 1\% in the area.
The attenuation effect was tested with a decrease of 30\% in signal amplitude.

\subsection{PMT characterization}
\label{sec:offline parameters}

A total of 422 PMTs were tested for DCR, gain, APP, and stability at room and cryogenic temperatures. 
All the measurements were performed using the experimental setup described in the previous section.
The same procedure was followed for each PMT to obtain comparable results.
A total of 422 PMTs were tested for DCR, gain, APP, and stability at room and cryogenic temperatures. All measurements employed the setup described earlier, with consistent procedures for comparable results. These parameters were selected as core criteria for their direct impacts on the liquid xenon dark matter detector’s performance: (1) Gain and associated Single Photon Electron (SPE) resolution govern PMT signal-to-noise ratio (SNR), critical for distinguishing weak S1/S2 signals from noise; (2) DCR mimics fake S1 signals (a major low-energy background source), demanding minimization for detection accuracy; (3) APP ensures signal fidelity, as after-pulses are easily confused with S2 signals and cause incorrect event reconstruction; (4) Stability monitoring targets light emission from PMT arcing, which disables PMTs and leads to data loss. While other relevant parameters, such as Quantum Efficiency (QE), Transit Time (TT), Transit Time Spread (TTS), and photocathode surface uniformity, are pre-specified and guaranteed by Hamamatsu, requiring no additional testing. The PMT dynamic range has been validated in our prior work~\cite{Luo:2023ebw}.

\subsubsection{PMT Gain}
The gain of a PMT, a critical parameter in light detection applications, is the ratio between the number of electrons in the PMT signal output at the anode and the number of photoelectrons emitted from the cathode.
The gain quantifies the amplification capability of the PMT to convert incoming photons into amplified electrical signals.
A higher gain corresponds to a higher signal-to-noise ratio, enabling the detection of weak light signals.

\begin{figure}[!htbp]
  \centering
  \includegraphics[scale=0.55]{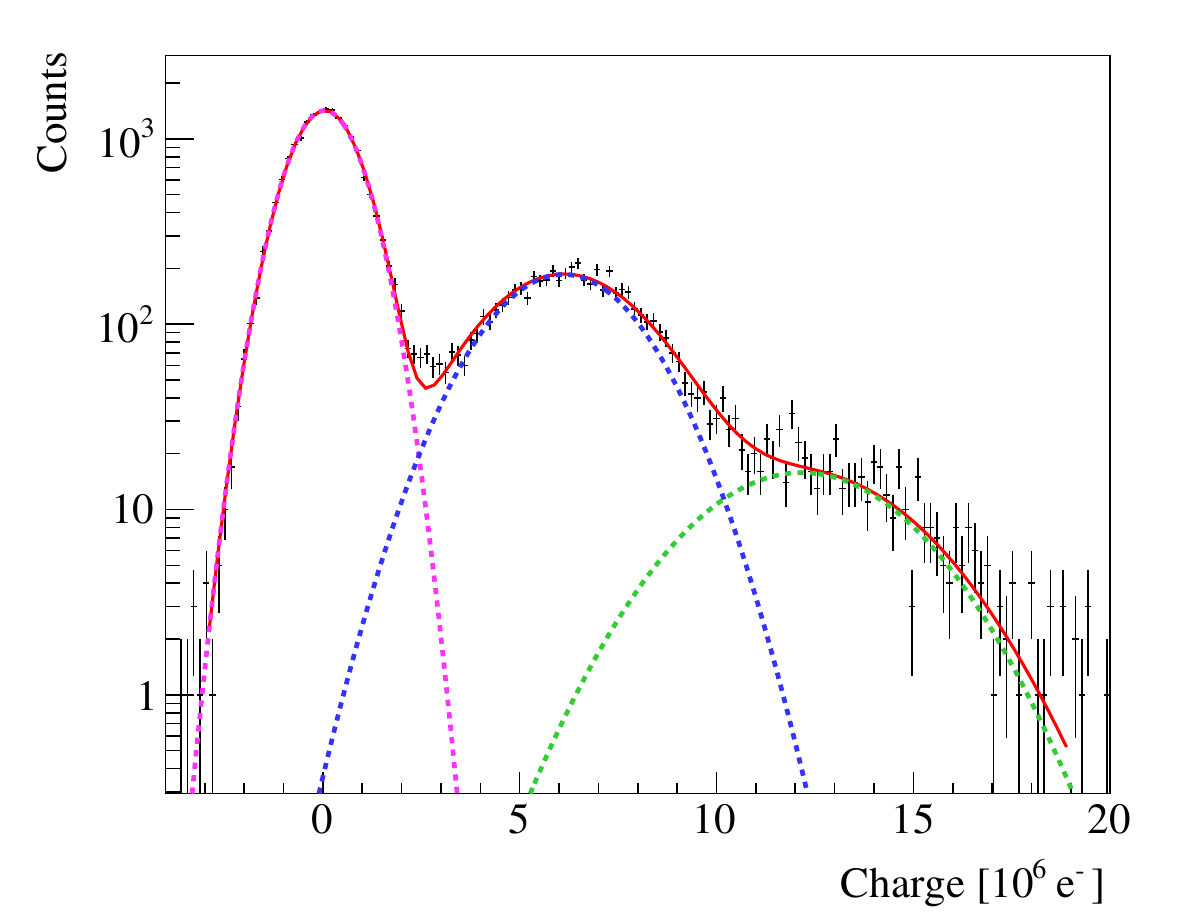}
  \caption{Typical PMT charge spectrum. The noise distribution is indicated by the pedestal (Magenta). The Gaussian curves of SPE and double PE peaks are blue and green, respectively. The red line represents the combined fit. }
  \label{fig:spe_gain_fit}
\end{figure}

The gain of the PMT was measured at a bias voltage of 1500~V. 
The LED light intensity was adjusted so that the PMTs would induce at most one or two photoelectrons per light pulse.
An external trigger synced to the LED light emission was used to acquire data in a time window of 800 ns.
The charge of each event was calculated from the sum of the ADC counts from the 40ns before and 40ns after the peak. 
A typical R11410-23 PMT charge spectrum is shown in Fig.~\ref{fig:spe_gain_fit}. 
The measured charge distribution of the PMT, denoted as $f(q)$, was fitted using the equation~\cite{Li:2015qhq}
\begin{equation}
\label{eq:fit_function}
 f(q) = c_0 \times G(q,\mu_0,\sigma_0) + c_1 \times G(q,\mu_1+\mu_0,\sqrt{(\sigma_1^2 + \sigma_0^2}) +c_2 \times G(q,2\mu_1+\mu_0, \sqrt{(2\sigma_1^2+\sigma_0^2)} \,
\end{equation} 
where $G(x, \mu, \sigma) = e^{(x-\mu)^2/2\sigma^2}$ denotes the standard Gaussian function. 
The first term in Eq.~\ref{eq:fit_function} describes the pedestal arising from the electronic noise.
The SPE resolution is defined as $\sigma_1/\mu_1$, where $\sigma_1$ and $\mu_1$ are the standard deviation and mean of the Gaussian function describing the SPE peak, respectively. 

\begin{figure*}[!htbp]
  \centering
  \includegraphics[scale=0.2]{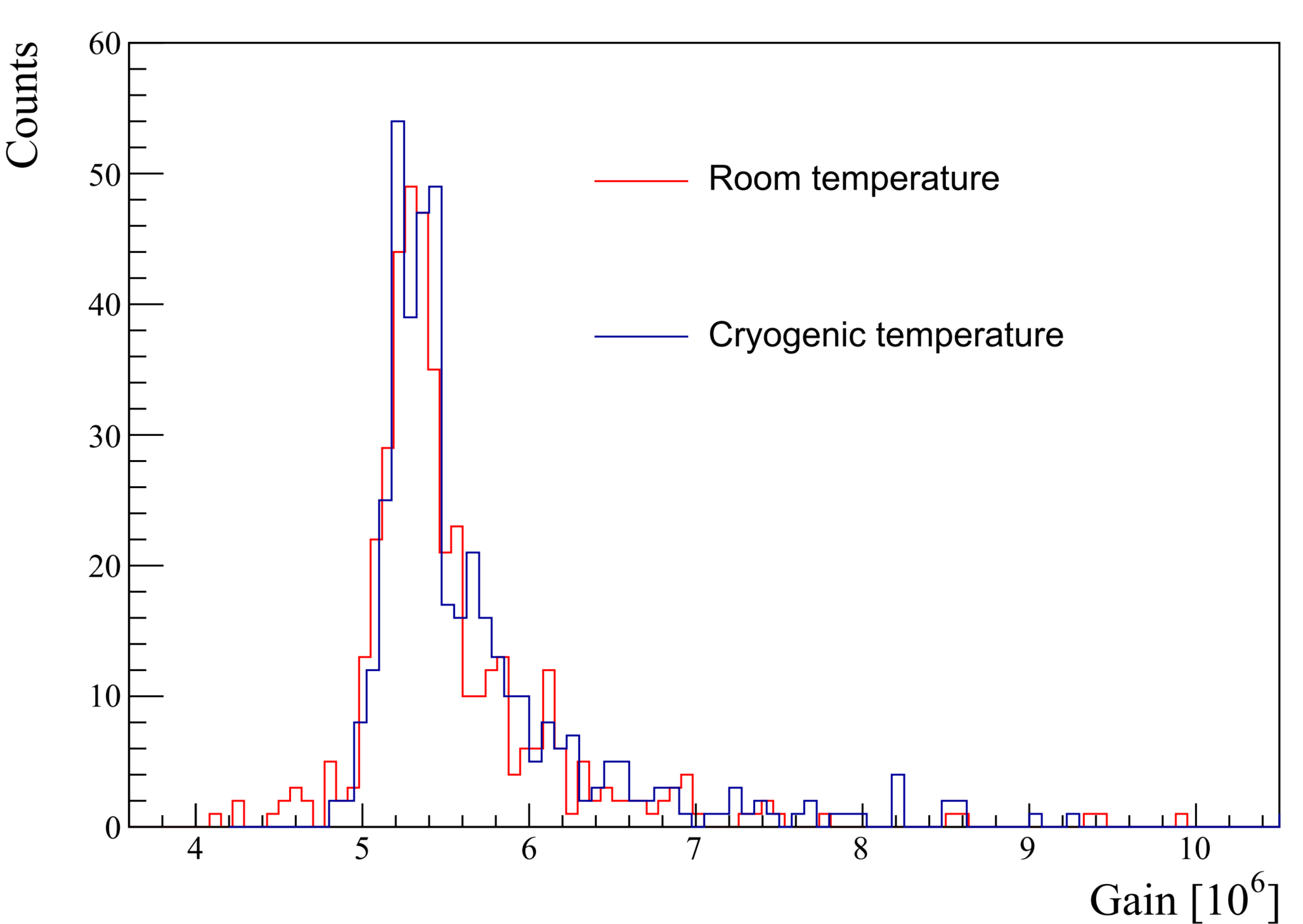}
  \caption{Gain distribution of all 422 PMTs at room (red) and cryogenic (blue) temperatures under 1500~V.}
  \label{fig:gain_hist}
\end{figure*}

Fig.~\ref{fig:gain_hist} shows the gain distribution of all 422 PMTs at room and cryogenic temperatures.
No obvious change was observed for the gain at the cryogenic temperature.
The average gain of all the PMTs was 5.5~$\times 10^6 $ (5.7~$\times 10^6 $) at room (cryogenic) temperature. The average SPE resolution is 28\% for both room and cryogenic temperatures.

\subsubsection{Dark count rate}
\begin{figure*}[htbp]
\centering
\includegraphics[scale=0.35]{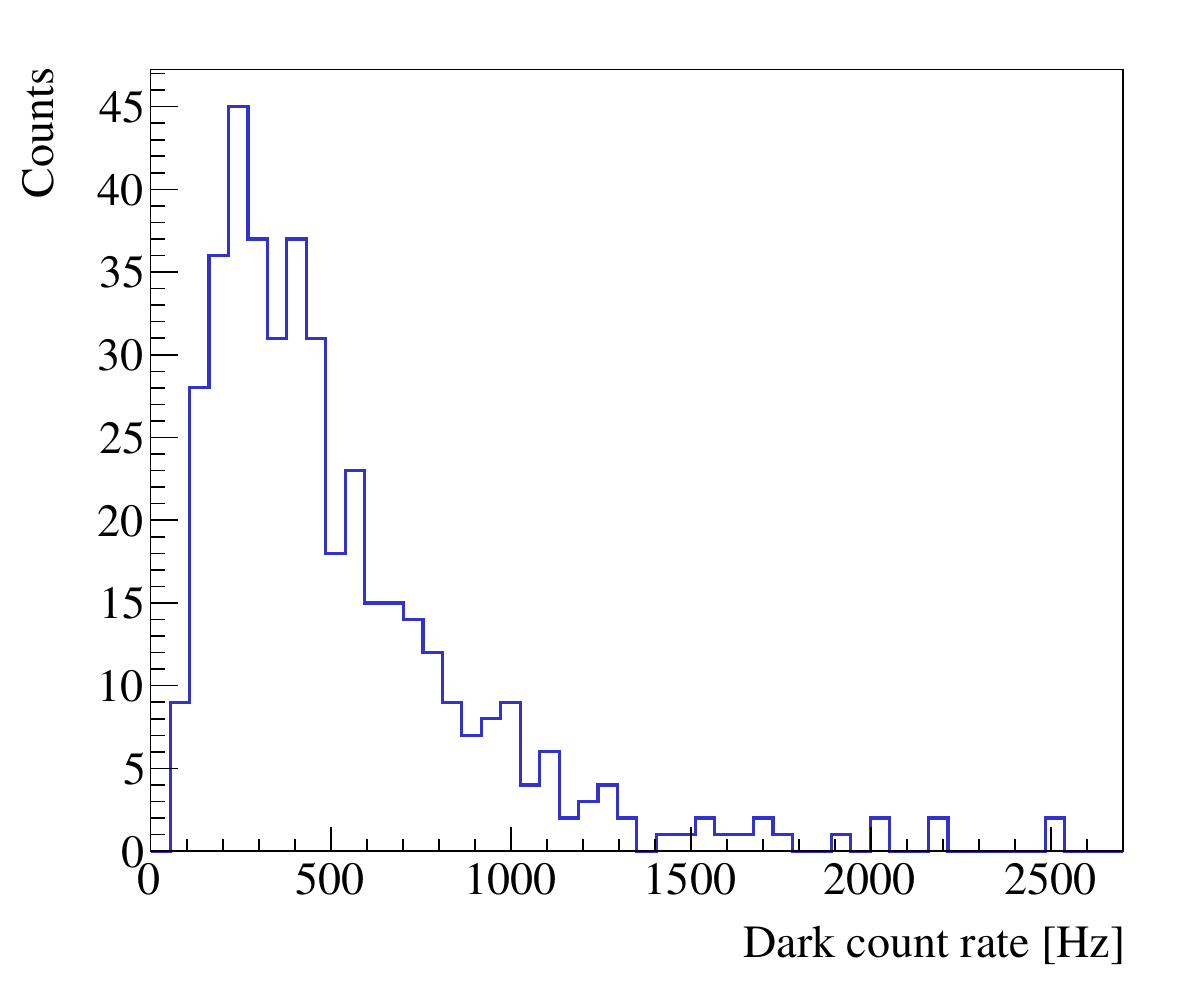}
\includegraphics[scale=0.35]{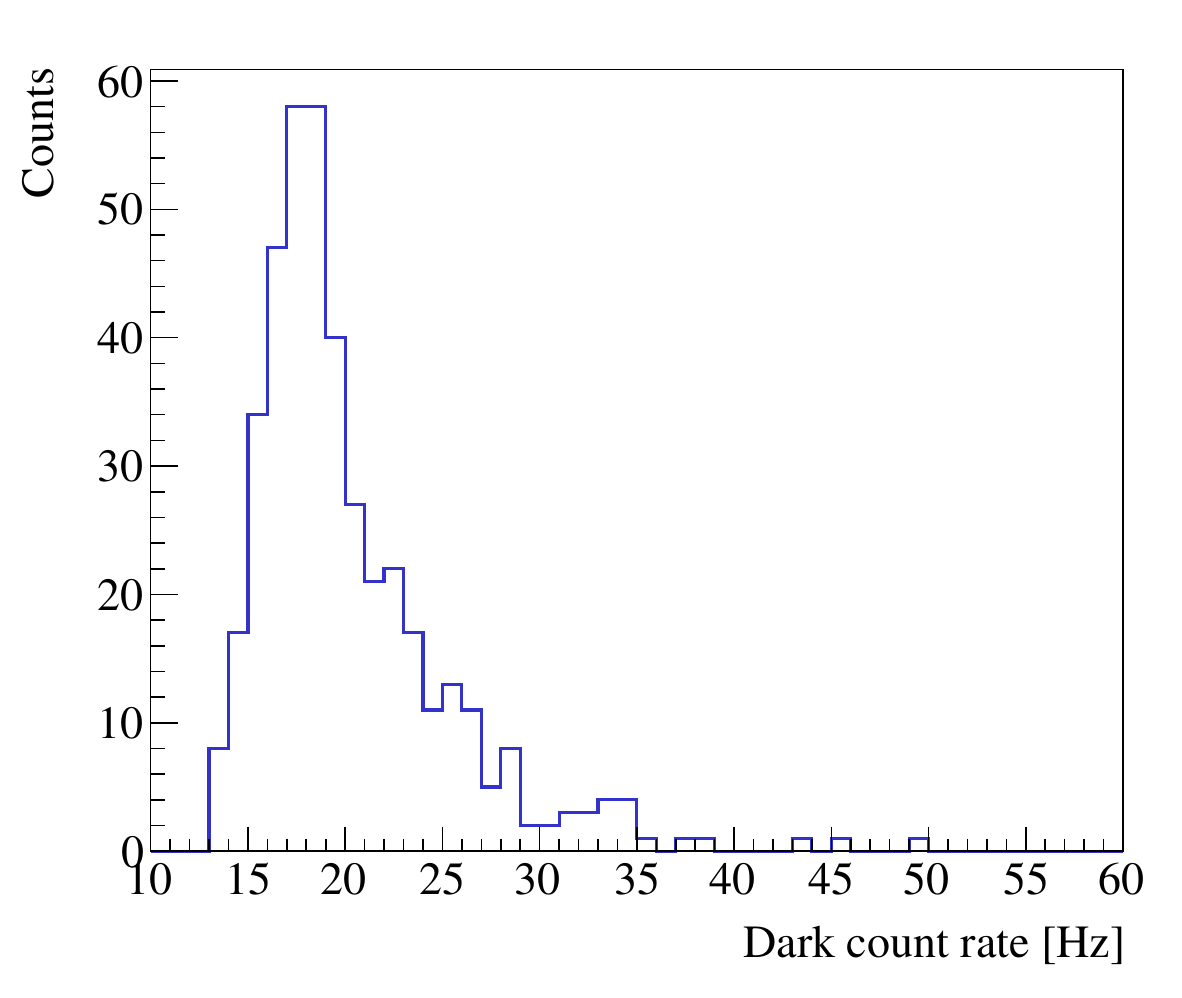}
\caption{Dark rate result at room and cryogenic temperatures.}
\label{fig:PMT_dr}
\end{figure*}

\begin{figure*}[tb]
  \centering
  \includegraphics[scale=0.4]{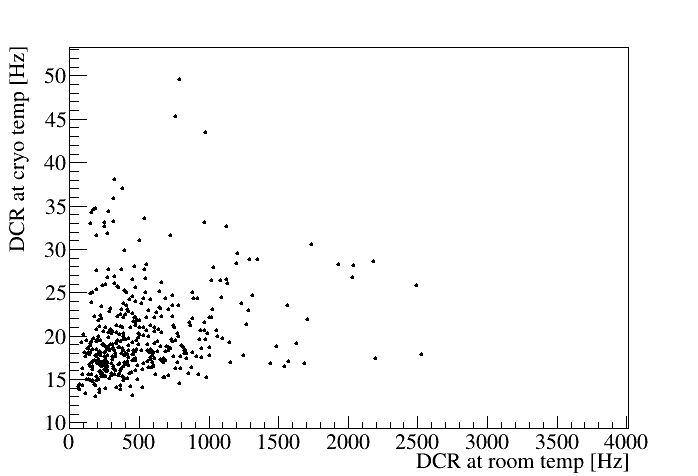}
  \caption{Correlation of DCR of PMTs at the room and cryogenic temperatures.}
  \label{fig:dcr_correlation}
\end{figure*}

The DCR is the rate at which the PMT generates signals without incident light. 
As mentioned earlier, a signal with a height greater than 20~ADC was recorded. 
The PMTs were placed in the light-tight Dewar for 8 hours and then tested over three thermal cycles. 
The thermal cycles simulated the temperature variations that the PMTs might encounter in the actual PandaX-4T experiment.
The DCR of PMTs, shown in Fig.~\ref{fig:PMT_dr}, varied between 50-2.5~kHz at room temperature and decreased to approximately 20~Hz at cryogenic temperature. 
This reduction is expected because a lower temperature reduces the thermal noise and leakage current from the photocathode, dynodes, and anode.
As shown in Fig.~\ref{fig:dcr_correlation}, it was calculated that the DCR correlation coefficient at room temperature and cryogenic temperature conditions was 0.1, indicating a relatively weak linear relationship.

In the DM search analysis, DCR contributes to the accidental background, whose magnitude is directly proportional to the number of PMTs and their respective DCRs. 
To mitigate this issue, we rejected PMTs with a DCR exceeding 20~kHz at room temperature or 200~Hz at cryogenic temperature.
This stringent selection criterion reduced the contribution of DCR-related backgrounds.

\subsubsection{After-pulse probability}
\label{sec:offline app}
We also measured the probability of unwanted after-pulses generated by ions from photoelectrons colliding with residual gases in the PMT vacuum.
Residual gases in the tube are inevitable due to either the desorption of the tube's components or gases (mainly helium) diffused through the PMT glass window.
When photoelectrons are accelerated, they may ionize the gas molecules.  
The resulting ions are then accelerated by the high voltage field in the reverse direction and hit the photocathode, which emits additional photoelectrons. 
These electrons are multiplied similarly to the original photoelectrons, resulting in a delayed pulse. 
Compared with the primary pulse, the after-pulse induced by ionization exhibits a signature time delay ranging from several hundred nanoseconds to several microseconds.
A high after-pulse rate in the PMTs can impair signal identification and energy resolution. 
Hence, PMT APP is another critical parameter that ensures high-quality data taking in PandaX-4T.

\begin{figure}[htbp]
\centering
\includegraphics[scale=0.4]{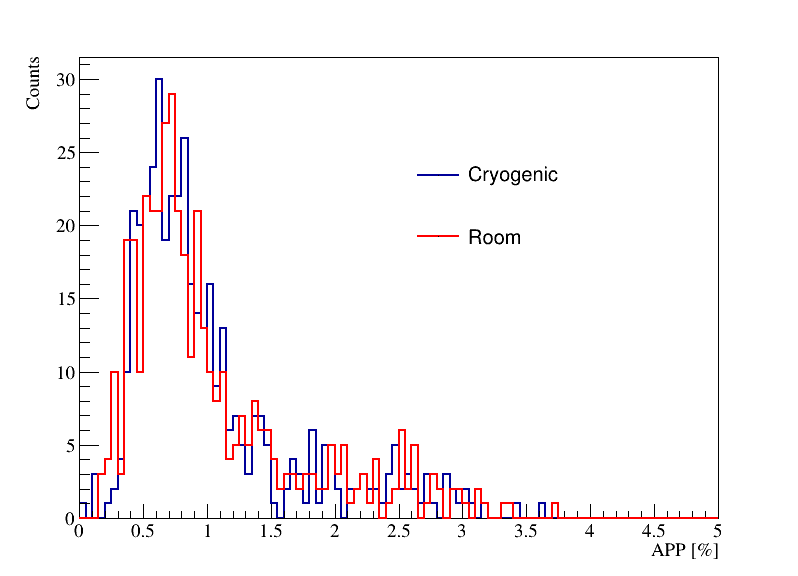}
\caption{APP result of tested PMTs at room and cryogenic temperatures.}
\label{fig:PMT_app}
\end{figure}

APP measurements were conducted at both room and cryogenic temperatures. 
After-pulses were defined as the small pulses from 0.2~$\mu$s to 5~$\mu$s window after the primary pulse.
APP was calculated as the ratio of the summed after-pulse count and the summed primary pulse charge.
The APP values in room and cryogenic temperatures are consistent and the distribution is shown in Fig.~\ref{fig:PMT_app}. 
The average APP is 1.04\% for both room and cryogenic temperatures. The acceptance threshold of the APP was determined to be 5\%. Consequently, none of the PMTs was rejected due to an elevated APP value.

\subsubsection{Spurious light emission}
Hamamatsu R11410 PMTs have been shown to emit light spuriously at both room and cryogenic temperatures~\cite{Akimov:2015cta,Li:2015qhq,Antochi:2021wik}. 
Although the exact cause of this process is unknown, it is widely presumed that the emission of single photons is from the PMT's internal structure.
Additionally, it is theorized that the residual alkaline material inside the PMT may be displaced due to thermal or physical shock, thereby disrupting the alkaline balance within the PMT. 
This imbalance in the alkaline material may lead to a small amount of light emissions. 
The random pairing of photons from PMT spurious light emission with single-electron signals in a TPC may imitate a dark matter signal. 
In PandaX-4T, this combination resulted in an important source of background.

\begin{figure}[tb]
  \centering
  \includegraphics[width=0.98\textwidth]{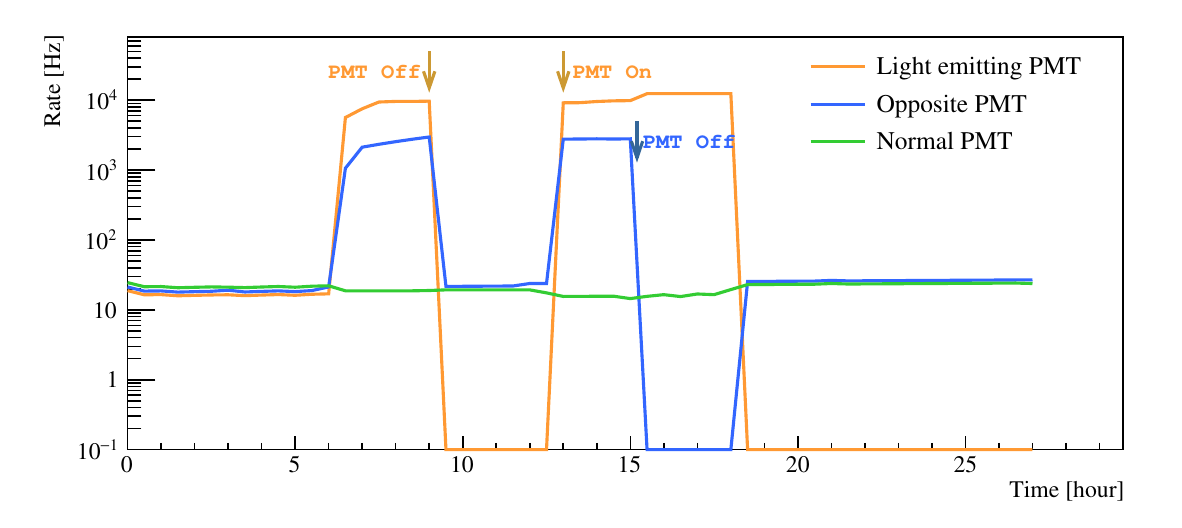}
  \caption{An example of light emission measurement, a clear correlation between the state of a light-emitting PMT and the DCR of the opposite PMT was observed. Normal PMT here refers to the PMT far away from the tested PMT.}
  \label{fig:light emission}
\end{figure}

In this test, PMTs were mounted in pairs facing each other that were only a few centimeters apart.
If one of the PMTs emitted photons, the opposite PMT would be able to detect an increasing change in the DCR. 
Fig.~\ref{fig:light emission} shows a clear correlation between the state of a light-emitting PMT and the hit rate of the opposite PMT.
When the test PMT was on, spurious photons were emitted by the test PMT, and the light signal seen by the opposite PMT increased from approximately 20~Hz to several thousand Hz. Normal PMT here refers to the PMT far away from the tested PMT.
The hit rate difference with the test PMT on and off defines the level of spurious light emission. 
Systematic tests show that the observed spurious light emission cannot be effectively eliminated by high voltage power cycling or bias voltage reduction. While rare room-to-cryogenic thermal cycles resulted in a temporary reduction in emission rate, this effect was non-reproducible and unstable across tests. To mitigate the associated background risk for the PandaX-4T experiment, all PMTs with abnormal spurious light emission were replaced by Hamamatsu.

\section{Performance in commissioning run}
\label{sec:commissioning run performance}
After testing, the PMTs were meticulously transported to the China JinPing Underground Laboratory (CJPL), which is the deepest underground laboratory in the world~\cite{CJPL}.  For enhanced safety and to preclude any potential damage, each PMT was individually encapsulated and stored within a nitrogen-filled bag. 
368 PMTs were mounted on the top and bottom of the PandaX-4T detector. Before installation, the PMTs, cables, and bases were meticulously cleaned using alcohol and pure water.
For the top (bottom) array, 169 (199) PMTs were arranged in a concentric circular (compact hexagonal) pattern. 
All PMTs were wrapped in Kapton film to insulate them from the supporting copper plate.
Each PMT and its corresponding base was securely held in position on the copper plate using three low-background stainless steel springs.
To increase the light collection efficiency, the front of the PMT array, facing the liquid xenon, was covered with PTFE reflectors.
To safeguard the PMTs, the grounded anode and bottom screen electrodes were positioned in front of the top and bottom arrays. 
There were 24 pieces of PTFE reflectors of 6~mm thickness surrounding the TPC's active volume. 
Additionally, 110 1~inch Hamamatsu R8520 PMTs were installed between the PTFE reflectors and the inner surface of the vessel.
Those PMTs were used to detect light signals outside the active volume for vetoing purposes.

Most PMTs worked well and the gain, DCR, and APP were stable during the PandaX-4T commissioning run. However, thirteen PMTs were broken, including four in the top array and nine in the bottom array.
Nine resulted from connection issues or damage to the high-voltage divider base.
The remaining four were turned off because of elevated APP.

\subsection{Gain stability}
The supplied bias voltages for PMTs were tuned in the range of 1400~V to 1550~V to obtain a uniform gain.
At the beginning of PandaX-4T data taking, the average gain was 5.7~$\times 10^{6}e^{-}$.
A weekly calibration with LED was utilized to monitor the gains.
We also extracted SPE spectra from physics data to continuously measure the PMT gains \emph{in situ}~\cite{PandaX:2023ggs}. 
The gains were stable during the data-taking campaigns, and the average value was 5.6~$\times 10^{6}e^{-}$ at the end of the first physics data-taking campaign.

A temporary gain drop was observed after calibrations with a strong kBq plutonium-carbon (PuC) neutron source.
After overnight calibration runs, we observed an average 5\% gain reduction in PMT gains.
The gain gradually recovered to its initial level within a week. 
We suspect that PMTs experienced a constantly high charge collection process due to high-frequency large light signals from neutron calibration, which temporarily affected the performance of the cathode and dynodes. 
Strong light exposure tests using LED were performed in an offline PMT setup.
The LED emitted light pulses at a frequency of 100~Hz and each PMT received a large number of photons with signals as large as $4\times 10^4$~PEs per pulse. 
The testing process lasted approximately 6~h and a similar PMT gain reduction was observed. 
Consequently, we concluded that the reduced PMT gain could be attributed to intense light exposure. 

\subsection{PMT DCR from random charges}
\begin{figure}[tbp]
  \centering
  \includegraphics[scale=0.25]{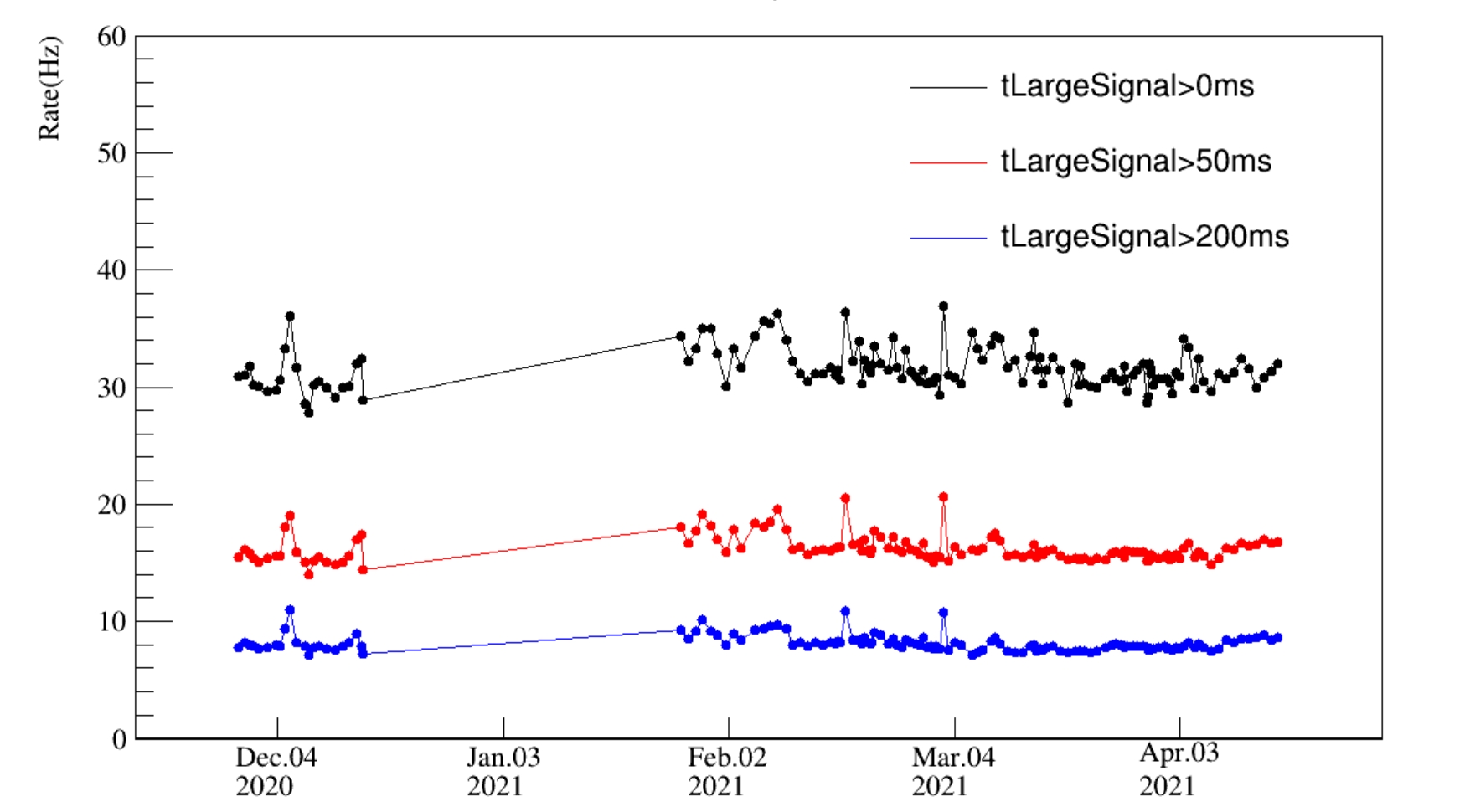}
  \caption{Random single charge evolution in PandaX-4T commissioning run. The random charge rate decreased with a larger delay time cut. The average random charge rate was 8~Hz with significantly delayed time greater than 200~ms.}
  \label{fig:dr evolution}
\end{figure}

When an event was recorded in PandaX-4T with only one PMT triggered, the charge of this PMT is defined as a new parameter called PMT random charge, which is a surrogate for PMT dark count.
For the physics data taking, the average recorded trigger rates of the top and bottom PMTs were around 1~kHz and 0.5~kHz, respectively, which included detector signals and dark counts of PMTs.
To select the dark counts of PMTs, a delay time cut was applied to ensure that the selected hit is far from a large signal~\cite{PandaX:2024med} (called tLargeSignal).
It ensured that any selected hit was sufficiently distant from a large signal (10000PE) to minimize the potential for signal contamination.
As shown in  Fig.~\ref{fig:dr evolution}, the average dark rate of all PMTs was calculated with different delay time cuts. The average random charge rate was 8~Hz with a significantly delayed time greater than 200~ms.
This value is later used to estimate the accidental background.  
The difference between the in-situ random charge rate and the offline cryogenic DCR measurement is attributed to the more stable liquid xenon environment in the underground detector, as well as the strict delay cut applied to in-situ data to exclude residual effects from high-energy detector events.

\subsection{After pulse evolution}
\label{sec:online app}
A detailed study of after-pulses in the R11410-23 was performed following the methods introduced in Ref.~\cite{Barrow:2016doe}. 
LED calibration was taken to monitor the APP before and after the data-taking campaign. The time distribution of after-pulses reflects their origin and mechanism. 

\begin{figure}[!htbp]
\centering
\subfigure[Charge versus delayed time of after pulses.]{
\label{fig:a}
\centering
\includegraphics[scale=0.2]{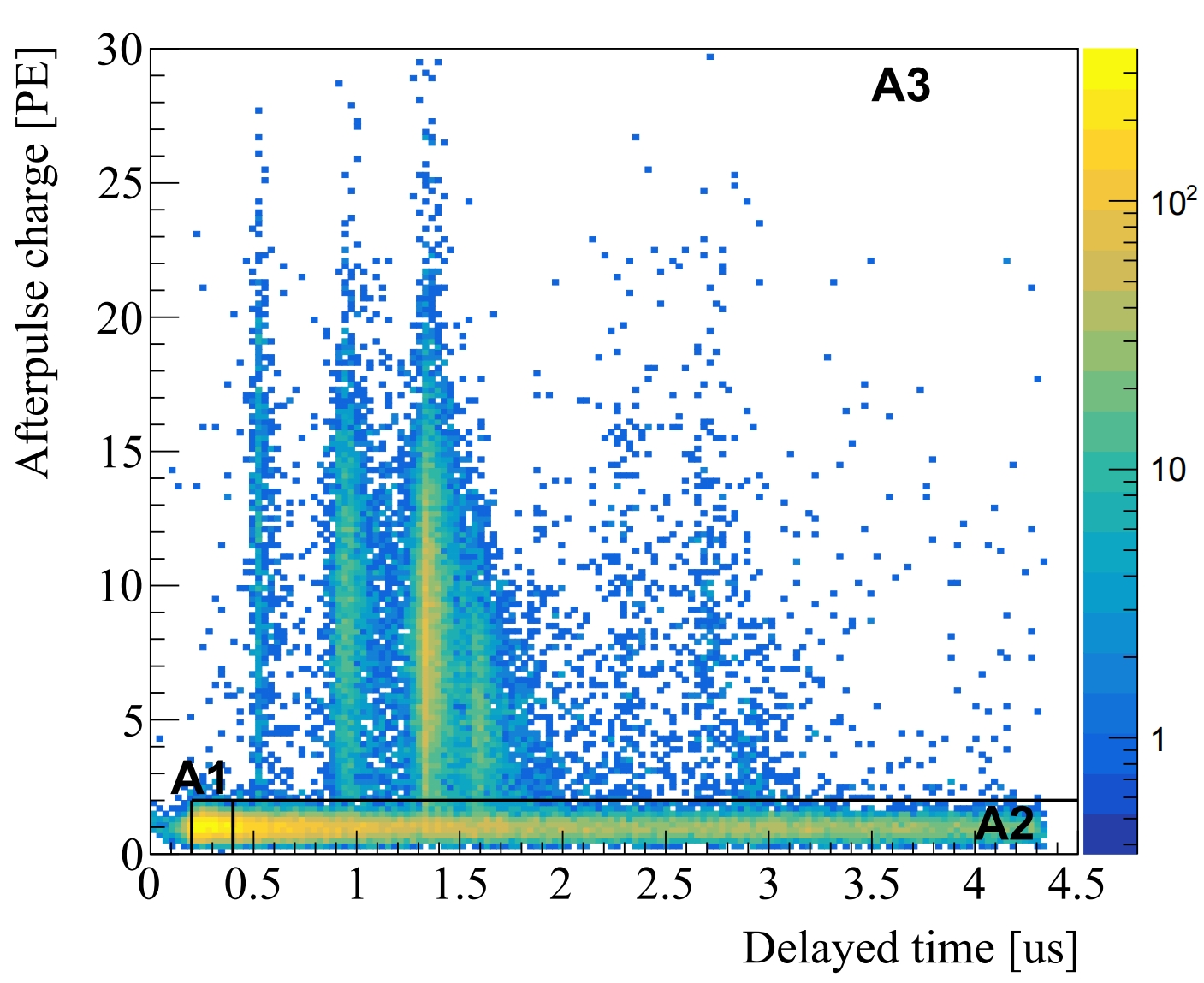}
}
\subfigure[Delayed time distribution.]{
\label{fig:b}
\centering
\includegraphics[scale=0.45]{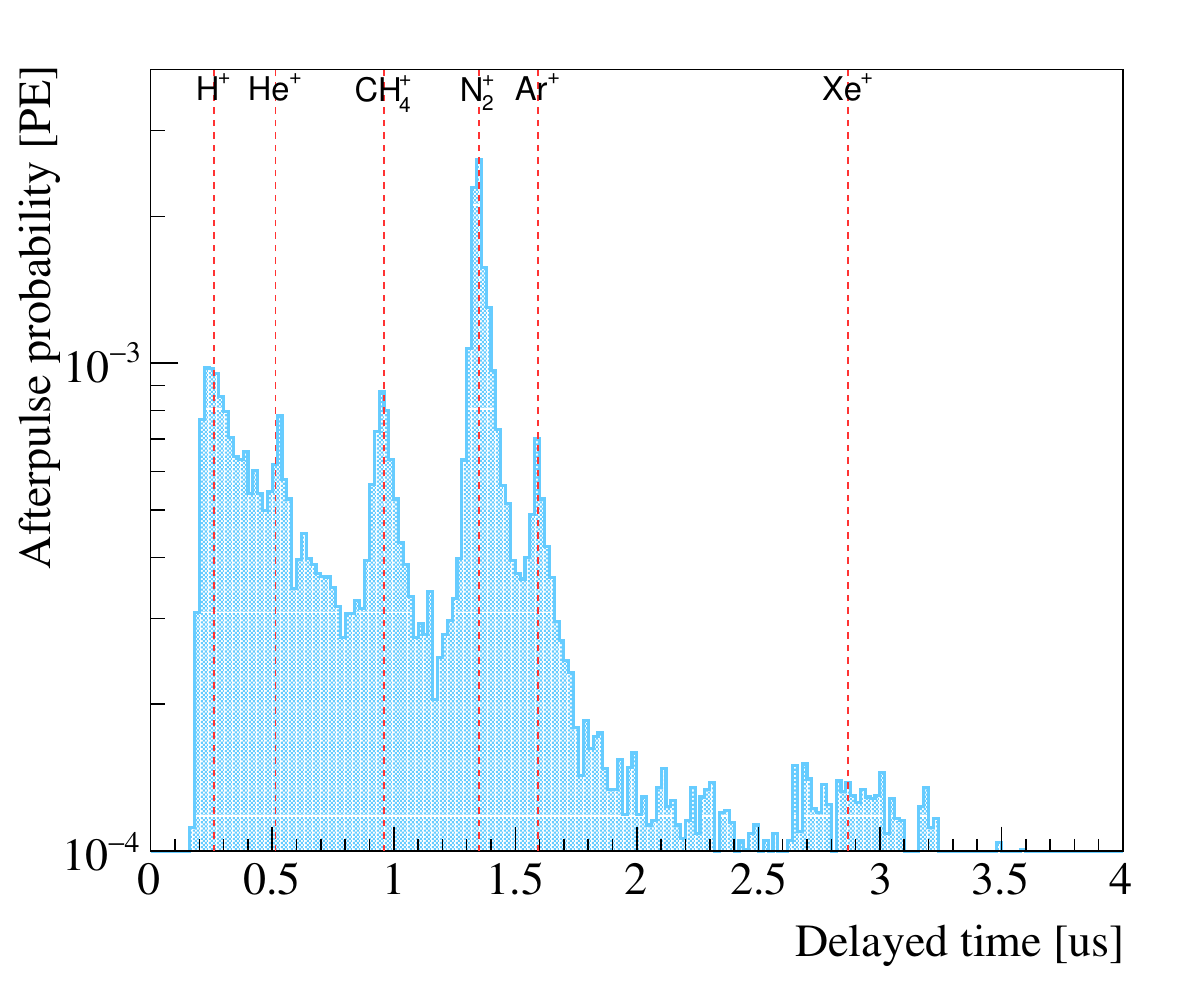}
}
\caption{Distributions of the after-pulses for an R11410-23 PMT with high afterpulsing rate: (a) charge
vs. delayed time, each point representing an after-pulse identified after the primary pulses; (b) the
delayed time distribution of the after-pulses in the A3 region with individual ionic components indicated.}
  \label{fig:afterpulse spectrum}
\end{figure}

In Fig.~\ref{fig:a}, A1 represents small pulses ($\textless$ 2~PE) that happened between 0.2~$\mu$s to 0.4~$\mu$s after the primary hit. The delay is mainly caused by
the elastic scattering of secondary electrons on the first dynode. A2 represents random hits ($\textless$ 2~PE) between 0.4 and 5~µs after the primary hit. A3 contains the after-pulses ($\textgreater$ 2~PE) caused by the residual or leaked gas ions. Peaks due to different ions can be identified from the timing distribution in Fig.~\ref{fig:b}. 
During the commissioning run period, the average APP is 1.35\%. It is slightly higher than that in offline tests, which can be attributed to the storage of the PMTs throughout one to two years before running.
During the experiment, we observed that a particular bottom PMT showed a high APP, exceeding 20\% . This exceptionally elevated APP value induced the PMT's incapacity to endure high voltages. 

\section{Summary}
\label{sec:summary}
We conducted a comprehensive characterization of 422 photomultiplier tubes for the PandaX-4T experiment. 
During this evaluation process, we meticulously examined parameters such as gain and dark count rate, with particular emphasis on addressing light emission and post-pulse issues. 
The manufacturer promptly replaced products failing to meet the specified standard. 

Key performance parameters were measured during the detector data taking.
Throughout the detector operation, a small fraction (5\%) of the photomultiplier tubes malfunctioned due to the PMT or the base.
The rest maintained constant gains over time.
The random charges and after pulses were under control as well.
The PMT arrays recently underwent an upgrade to replace the bases for a wider dynamic range, at the same time, we reduced the number of shared NHV channels from eight to four. 
More physics results will be expected with the new data-taking campaigns of PandaX-4T.

\acknowledgments
This work has been supported by the Ministry of Science and Technology of China (No. 2023YFA1606203 and 2023YFA1606204), the National Natural Science Foundation of China (No. 12222505), Shanghai Pilot Program for Basic Research-Shanghai Jiao Tong University (No. 21TQ1400218), National Science Foundation of Sichuan Province (No. 2024NSFC1371). We also thank the sponsorship from the Chinese Academy of Sciences Center for Excellence in Particle Physics (CCEPP), Hongwen Foundation in Hong Kong, New Cornerstone Science Foundation, Tencent Foundation in China, and Yangyang Development Fund.

\bibliography{appendix.bib}
\bibliographystyle{elsarticle-num}

\nolinenumbers
\end{document}